\documentclass[a4paper,11pt]{article}
\usepackage{jheppub} % for details on the use of the package, please
\usepackage[T1]{fontenc} % if needed

\usepackage{makeidx} % MakeIndex
\makeindex % MakeIndex
\usepackage{datetime}
\vbadness=\maxdimen

\usepackage{CJKutf8}
\newenvironment{korean}
{\begin{CJK}{UTF8}{}\CJKfamily{mj}}
{\end{CJK}}

\usepackage{amstext}
\usepackage{amsfonts}
\usepackage{amsmath}
\usepackage{amssymb}

\usepackage{bbding} 
\usepackage{cancel}
\usepackage{color}
\usepackage{empheq} % box package

\usepackage{enumitem}
\usepackage{fancyvrb} % writing down the source codes
\usepackage{float} % figure location fixing. [H] is possibe.
\usepackage{footnote} % footnote in table
\makesavenoteenv{tabular} % footnote in table
\makesavenoteenv{table} % footnote in table
\usepackage{hhline} %double line in the table
\usepackage{hyperref} % to use URL link
\usepackage{makecell}
\usepackage{mathrsfs}
\usepackage{mathtools} % make possible to add a long underbrace 
\usepackage{multirow}
\usepackage{pifont}
\usepackage{simplewick} % Wick contraction (http://tug.ctan.org/macros/latex/contrib/simplewick/simplewick.pdf)
\usepackage{slashed} % Dirac slash
\usepackage{stackrel}
\usepackage{tipa} % lambdabar notation
\usepackage[normalem]{ulem}
\usepackage{varioref} % reference style
\usepackage{wasysym} % more symbols
\usepackage[dvipsnames]{xcolor} % including color more

\usepackage{graphicx}
\allowdisplaybreaks
\usepackage[normalem]{ulem}

\title{\boldmath 
Exact Greybody Factors for the Brane Scalar Field\\[2mm] of Five-dimensional Rotating Black Holes}

\author[a]{Young-Hwan Hyun}
\author[b]{, Yoonbai Kim}
\author[c]{, Seong Chan Park}
\affiliation[a]{Korea Institute of Science and Technology Information (KISTI), 245 Daehak-ro, Yuseong-gu, Daejeon 34141, Republic of Korea}
\affiliation[b]{Department of Physics and Institute of Basic Science,
Sungkyunkwan University, Suwon 16419, Republic of Korea}
\affiliation[c]{Department of Physics and IPAP, Yonsei University, Seoul 03722, Republic of Korea}
\emailAdd{yoonbai@skku.edu}
\emailAdd{sc.park@yonsei.ac.kr}
\abstract{
We study  scalar perturbations of the five-dimensional rotating black holes and find an exact solution giving exact description of the Hawking radiation.
Mathematically, the full solution for this $s$-wave mode is expressed
in terms of the prolate spheroidal wave function with complex parameters.
By using the spheroidal joining factor, we write the corresponding boundary condition and greybody factors. We also check that the exact result reproduces the low frequency limit of the greybody factor and shows good agreement with the known numerical results.}

\begin{document}

%\today
\maketitle
\flushbottom

\section{Introduction}

Higher dimensional black holes have been attractive objects 
which may shed light on understanding extra dimensions.
A particular interest has been received  for the low-scale gravity scenario~\cite{ArkaniHamed:1998rs,Antoniadis:1998ig,ArkaniHamed:1998nn,Randall:1999ee,Randall:1999vf} due to a possibility  to be produced in proton-proton collisions at the LHC~\cite{Giddings:2001bu,Dimopoulos:2001hw} and comic ray-nucleon collisions at the atmosphere of the earth~\cite{Anchordoqui:2001cg,Feng:2001ib,Ringwald:2001vk,Jho:2018dvt}. (See the comprehensive reviews~\cite{Park:2012fe,Kanti:2004nr,Cavaglia:2002si}.)

In the low energy gravity scenarios, the fundamental gravity scale is as low as the electroweak scale, $M_* \sim {\cal O}(1)$ TeV, and thus
the trans-Planckian particle collision with the collision energy, $\sqrt{s} \gg M_*$, generates 
higher dimensional black holes with a mass larger than the Planck scale, $M_{\rm bh} \gg M_*$.
In this paradigm, classical laws of black hole physics still hold as long as the black hole is larger than the Planck length, $r_{\rm bh} \gg 1/M_*$. 
After the production of the black holes with a mass larger than the Planck scale,
it is anticipated to decay via emitting Hawking radiation~\cite{Hawking:1974sw}.
During the spin-down and Schwarzschild phase, 
the produced black hole radiates particles and loses its mass 
until the mass becomes of Planck scale where the Hawking radiation is not applicable in the present form.
This process allows to test whether the extra-dimensions exist
or not and to see which low scale gravity model gives us good approximations.
For this reason, studying the higher dimensional, rotating black holes in detail
is getting important.\\

To identify the signal of any higher dimensional black hole,
the study on the propagation of the particles from the black holes must be preceded.
According to some of the low scale gravity scenarios~\cite{ArkaniHamed:1998rs,Antoniadis:1998ig,ArkaniHamed:1998nn,Randall:1999ee,Randall:1999vf}, gravitons propagate in the bulk, while the standard model particles are confined to live 
on a $(1+3)$-dimensional brane. The black holes radiate mainly on the brane~\cite{Emparan:2000rs}. 
The propagation of these perturbations is governed by the Teukolsky equation~\cite{Teukolsky} in the background of a higher-dimensional, rotating black hole. Once the Teukolsky equation is given~\cite{Ida:2002ez}, a cornerstone is to find the exact solution which enables us various relevant analyses including the the Hawking radiation and the stability of the black holes. However, finding an exact solution of the Teukolsky equation is tough and thus previous investigations assumed a low-frequency regime~\cite{Ida:2002ez,Creek:2007sy,Creek:2007tw,Jorge:2014kra} or only used numerical methods~\cite{Ida:2005ax,Ida:2006tf, Harris:2005jx,Duffy:2005ns,Casals:2005sa,Casals:2006xp}. 
Overcoming these restrictions and finding the analytic, exact expressions for the greybody factor for the higher dimensional, rotating black hole  is extremely important but never has been successful. Only limited results for the non-rotating black hole are found for spherically symmetric, static black hole in $d$ dimension~\cite{Harmark:2007jy}.  In this work, we could make some partial success in finding the exact solutions: the fully analytic results for the scalar mode on the brane without the low frequency restriction are found! 
Previously known analytic and numerical results are successfully re-driven from our results.

This paper is composed as follows: in section~\ref{Sec2}, we find the general $s=0$ solution of the Teukolsky equation on the brane in an exact form. In section~\ref{Sec3}, the boundary conditions are assigned, appropriate for the ingoing waves. In section~\ref{Sec4}, the exact ingoing wave solution is applied to compute the greybody factors and the obtained exact formulas are compared to the approximate results. We conclude the section~\ref{Sec5} with discussions. 
\vspace{12pt}

\section{Exact solutions of generalized Teukolsky equations}
\label{Sec2}

The generalized Teukolsky equation for the brane fields 
in the 5D spacetime is obtained by applying the Newman-Penrose formalism 
to the four-dimensional induced metric~\cite{Ida:2002ez}.
Then the perturbation equations are given as separated differential equations 
with a separation constant $A$, 
\begin{align}\label{A-TeuEq}
&\frac{1}{\sin \theta}\frac{{\textrm{d}}}{{\textrm{d}}\theta}
        \left(\sin\theta\frac{{\textrm{d}}S}{{\textrm{d}}\theta}\right)
        +[(s-a \omega \cos\theta)^{2}
        -(s\cot\theta+m\csc\theta)^{2}-s(s-1)+A]S=0,
        \\
&\Delta^{-s}\frac{{\textrm{d}}}{{\textrm{d}}r}
        \left(\Delta^{s+1}\frac{{\textrm{d}}R}{{\textrm{d}}r}\right)
        +\bigg[
        \frac{K^{2}}{\Delta}
        +2irs \bigg(2\omega-\frac{K}{\Delta}\bigg)
        +2ma\omega-a^{2}\omega^{2}-A
        \bigg] R=0, 
\label{R-TeuEq}
\end{align}
where $s$ is the spin-weight of the fields, 
$\omega$ is a frequency of the fields, and $K(r)=(r^{2}+a^{2})\omega-ma$.

The solution to the 5D angular Teukolsky equation~\eqref{A-TeuEq} 
is given by a generalized spheroidal function, 
and the separation constant is determined 
in terms of the spin-weighted spheroidal eigenvalue 
${}_{s}\lambda_{l}^{m}(\pm ia\omega)$ 
which reduces to a spheroidal eigenvalue for $s=0$ case~\cite{Frolov:2002xf,Ida:2002ez}, 
\begin{eqnarray}\label{A}
A = {}_{s}\lambda_{l}^{m(a)}(\pm i a \omega) - a^{2}\omega^{2} - s(s+1),
\end{eqnarray}
where the superscript ${}^{(a)}$ denotes the eigenvalue for the angular equation \eqref{A-TeuEq}.
With this value $A$, all parameters in the radial equation are fixed.

The 5D radial Teukolsky equation \eqref{R-TeuEq} possesses 
two regular singularities in the finite region 
and one irregular singularity of rank $1$ at infinity. 
Thus the equation transforms to the confluent Heun equation 
which is the second order linear differential equation 
with the same singularity structure. 
After changing variables to dimensionless ones, 
$a \rightarrow a_{*} \equiv a/r_{{\rm h}}$, 
$\omega \rightarrow {\tilde \omega} \equiv r_{{\rm h}} \omega$, and 
$r \rightarrow {\tilde r} \equiv r / r_{{\rm h}}$, 
the transformation of the dependent variable, 
$R(\tilde r)=({\tilde r}+1)^{-\frac{s}{2}}({\tilde r}-1)^{-\frac{s}{2}}{\tilde R}(\tilde r)$, 
leads to the confluent Heun equation in the B\^ocher symmetrical form, 
so-called the general spheroidal equation~\cite{Ronveaux},
\begin{align}
\frac{{\textrm{d}}}{{\textrm{d}} {\tilde r}} (1-{\tilde r}^{2}) 
\frac{{\textrm{d}}}{{\textrm{d}}{\tilde r}} {\tilde R}({\tilde r}) 
+ \Big[
\gamma^{2}(1-{\tilde r}^{2})
-2 i \gamma \beta {\tilde r} + \lambda 
- \frac{\mu^{2}+\sigma^{2}+2\mu \sigma {\tilde r}}{1-{\tilde r}^{2}}
\Big] 
{\tilde R}({\tilde r}) = 0,
\label{GSE}
\end{align}
where we find a set of parameters,
\begin{align}\label{para_rel}
\begin{cases}
\gamma={\tilde \omega} \\
\mu =i[(a_{*}^{2}+1){\tilde \omega}-ma_{*}] \\
\lambda^{\mu}_{\nu}(\gamma)=A-(a_{*}^{2}+2) {\tilde \omega}^{2} + s(s+1) \\
\beta=\sigma = s \\
\end{cases}
.
\end{align}
With this set of parameters we can write down the solution in terms of the confluent Heun functions with the general complex parameters, which has not been studied much. In the case of the scalar fields ($s=0$), the parameters $\sigma$ and $\beta$ become zero, and the equation~\eqref{GSE} reduces to the spheroidal equation with general parameters developed by Meixner~\cite{Meixner1, Meixner2} and explained well with calculation methods in Ref.~\cite{Falloon}. Therefore, we focus on the $s=0$ case in this paper, in which the Teukolsky equation becomes
\begin{align}
\label{s=05DTeuEq}
&\frac{1}{\sin\theta}\frac{{\rm d}  }{{\rm d} \theta}\left(\sin\theta\frac{{\rm d} }{{\rm d} \theta}S(\theta)\right)+[(a \omega \cos\theta)^{2}
        -(m\csc\theta)^{2}\textcolor{magenta}{}+A]S(\theta)=0\,,\nonumber\\ 
&\frac{{\rm d}  }{{\rm d}r}\left(\Delta\frac{{\rm d}  }{{\rm d} r}R(r)\right)+\bigg[
        \frac{K^{2}}{\Delta}
        +2ma\omega-a^{2}\omega^{2}-A
        \bigg] R(r)=0\,,
\end{align}
whose systematic derivation with uniqueness is discussed in Appendix~\ref{AppB}.

In the spheroidal equation\footnote{For the spheroidal equation and the related mathematical properties, refer to Appendix~\ref{AppC}.}, corresponding to the scalar case, 
three parameters $\mu$, $\nu$, and $\gamma$ are called 
the order, degree, and size parameters, respectively. 
For the real (purely imaginary) size parameter $\gamma$, 
the equation is called the prolate (oblate) spheroidal equation.
A constant $\lambda_{\nu}^{\mu}(\gamma)$ in~\eqref{para_rel} is called the spheroidal eigenvalue 
and is determined as the minimal solution~\cite{Falloon}. 
In our case, the eigenvalue $\lambda$ is fixed 
by the relation with the angular Teukolsky equation~(\ref{para_rel}) 
and so does $\nu$. 
The parameter domain is given by
\begin{align}
\mu, \nu \in \mathbb{C},
\qquad \gamma \in \mathbb{R^{+}}, 
\qquad {\tilde r}>1~. 
\end{align}
Therefore the solutions for the 5D Teukolsky equation for the scalar field 
are given by the radial prolate spheroidal functions 
$S_{\nu}^{\mu}({\tilde r;\gamma})$ $(i=1,2)$ with complex parameters, 
where the radial in the radial spheroidal function indicates 
the domain ${\tilde r}>1$ of the spheroidal function, 
not the radial Teukolsky equation. 
Since the radial spheroidal function 
$S_{\nu}^{\mu}({\tilde r};{ \gamma} )$ involves branch cuts 
in the complex ${\tilde r}$-plane along the semi-infinite line 
starting at the point ${\tilde r}=0$ and passing through ${\tilde r}=-1/{\gamma}$ 
for non-integer $\nu$, 
and on the interval $(-1,1)$ for non-integer $\mu/2$~\cite{Falloon}, 
the branch cut does not lie on the real axis ${\tilde r}>1$ of our consideration. \\

%\section{Full solutions and the ingoing boundary condition  for scalar fields}

Since the Wronskian of the radial spheroidal function is proportional to 
$[{\tilde \omega}({\tilde r}^{2}-1)]^{-1}$ 
for all parameter values~\cite{Meixner1,Falloon}, 
the full solution and its eigenvalues of the 5D generalized Teukolsky equation 
for the brane scalar fields are given by,
\begin{align}\label{FullSol_SS}
\text{solution : }& 
        R({\tilde r }) 
        = C_{1}S_{\nu}^{\mu(1)}({\tilde r };{\tilde \omega })
        + C_{2}S_{\nu}^{\mu(2)}({\tilde r };{\tilde \omega })\,, 
        \nonumber\\
\text{eigenvalues : }& 
        \lambda_{\nu}^{\mu}({\tilde \omega }) 
        = \lambda_{l}^{m (a)}(i a_{*} {\tilde \omega }) 
        - 2(a_{*}^{2} + 1){\tilde \omega }^{2}\,,
\end{align}
where $C_{1}$ and $C_{2}$ are constants, and $\mu=i[(a_{*}^{2}+1){\tilde \omega }-ma_{*}] $.
Since the spheroidal eigenvalue has the symmetry property, \begin{align}
\label{lambdasym}
\lambda_{\nu}^{\mu}({\tilde \omega})
= \lambda_{-\nu-1}^{\mu}({\tilde \omega})
= \lambda_{\nu}^{-\mu}({\tilde \omega})\,,
\end{align}
we choose one of the three parameter sets, $(\mu,\nu)$, $(\mu,-\nu-1)$, and $(-\mu,\nu)$, and then keep the calculation. The linearly independent solutions $S_{\nu}^{\mu(i)}({\tilde r };{\tilde \omega })$ can be expanded in terms of the spherical Bessel functions, $j$, $y$,
\begin{align}\label{Sdef}
S^{\mu\ (i)}_{\nu}({\tilde r };{\tilde \omega}) 
        = \frac{(1-1/{\tilde r }^{2})^{\mu/2}}
               {A^{-\mu}_{\nu}({\tilde \omega})}
        \sum_{k=-\infty}^{\infty} 
                a^{-\mu}_{\nu,2k}({\tilde \omega})
                f_{\nu+2k}({\tilde \omega}{\tilde r })~,
\end{align}
where $i=1,2$, and $f=j,y$, respectively, 
and the normalization factor $A^{\mu}_{\nu}({\tilde \omega})$ 
is given by 
\begin{align}\label{Norm_factor}
A^{\mu}_{\nu}({\tilde \omega}) 
        = \sum_{k=-\infty}^{\infty} 
        (-1)^{k} a^{\mu}_{\nu,2k}({\tilde \omega})~.
\end{align}
As ${\tilde \omega} {\tilde r} \rightarrow \infty$, the factor~\eqref{Sdef} is chosen so that the radial functions have the following asymptotic behaviours~\cite{Falloon}:
\begin{align}\label{SS_jy_sc}
S^{\mu (1)}_{\nu} ({\tilde r};{\tilde \omega} ) 
\longrightarrow j_{\nu}({\tilde \omega} {\tilde r}),\qquad
S^{\mu (2)}_{\nu} ({\tilde r};{\tilde \omega} ) 
\longrightarrow y_{\nu}({\tilde \omega}{\tilde r})\,.
\end{align}
In the full solution, 
the series coefficients $a^{\mu}_{\nu,2k}({\tilde r };{\tilde \omega })$
satisfy the three-term recurrence relation,
and the solutions to the recurrence relations are given 
as the minimal solution with which the ratio of the coefficients 
converges to zero, 
$a^{\mu}_{\nu,2k\pm2}({\tilde \omega })
/a^{\mu}_{\nu,2k}({\tilde \omega}) \rightarrow 0$, 
as $k \rightarrow \pm \infty$. 
This minimal condition is satisfied
for the spheroidal eigenvalues $\lambda_{\nu}^{\mu}({\tilde \omega})$, 
and the series expansion in the definition absolutely converges 
for all ${\tilde \omega}$ and ${\tilde r}$. 
\vspace{12pt}

\section{Boundary condition near the horizon}
\label{Sec3}
Near the horizon of a black hole, the coefficient of the wave solution corresponding to outgoing wave packet should vanish due to its attractive nature and thus this is called ingoing boundary condition. 
The asymptotic near-horizon and far-field solutions 
for the generalized Teukolsky equation~\eqref{R-TeuEq} were obtained 
in the previous research~\cite{Ida:2002ez} 
\begin{center}
\begin{tabular}{|cc|cc|}\hline
\multicolumn{2}{|c|}{$r \rightarrow \infty$} 
        & \multicolumn{2}{|c|}{$r \rightarrow r_{{\textrm{h}}}$} \\\hline\hline
~~~ingoing~~~ & ~~~outgoing~~~ & ingoing & outgoing \\\hline
$\displaystyle{\frac{1}{r} e^{-i \omega r_{*}}} $ 
        & $\displaystyle{\frac{1}{r^{2s+1}} e^{i \omega r_{*}}} $
        & $e^{-i k r_{*}}\Delta^{-s} $
        & $e^{i k r_{*}}$ \\
$\displaystyle{~\rightarrow \frac{1}{r} e^{-i \omega r}}$
        & $\displaystyle{~\rightarrow \frac{1}{r} e^{i \omega r}}$
        & $\rightarrow \left( {\tilde r}-1 \right)^{- \frac{\mu}{2}}$
        &$ \rightarrow \left( {\tilde r}-1 \right)^{ \frac{\mu}{2}}$ 
        \\\hline
\end{tabular}
\end{center}
where the tortoise coordinate $r_{*}$ was used, $k=\omega-ma/(r_{{\textrm{h}}}^{2}+a^{2})$, and $dr_{*}/{dr}={(r^{2}+a^{2})}/{\Delta(r)}$.
By definition, the coordinate $r_{*}$ approaches $r$, $r_{*} \rightarrow r$ for $r \rightarrow \infty$. 
In the last row of the table, the asymptotic solutions are written 
in the Boyer-Lindquist coordinates. 
To identify the ingoing boundary condition, 
we need to find two linear combinations of the two solutions 
$S_{\nu}^{\mu(i)}({\tilde r };{\tilde \omega })~(i=1,2)$ 
in order that the two resultant functions become either ingoing or outgoing solution in the near horizon limit, respectively. Overall factors of the chosen solutions $S_{\nu}^{\mu(i)}({\tilde r };{\tilde \omega })~(i=1,2)$ in~\eqref{Sdef} match the outgoing asymptotic solution in the near-horizon limit in the table, however there remain nontrivial series terms in the solution to be examined. Therefore, a different linear combination matching the ingoing asymptotic solution should be investigated in what follows by using the properties of  the spheroidal functions, which are nontrivial.\\

 The spheroidal functions have joining relations 
between the second kind angular spheroidal function of type II, 
$\text{Qs}^{\mu}_{\nu}({\tilde r };{\tilde \omega })$, 
which is expanded in terms of the associated Legendre functions of type II, 
$\mathfrak{Q}^{\mu}_{\nu}({\tilde r };{\tilde \omega })$, 
and the radial spheroidal function, 
$S^{\mu (1)}_{\nu} ({\tilde r };{\tilde \omega })$ 
as\footnote{In the reference~\cite{Falloon}, 
the relation for $S^{\mu(2)}_{\nu}$ is given with wrong signs.},
\begin{align}
\label{K_def}
&S^{\mu (1)}_{\nu} ({\tilde r };{\tilde \omega })
= K^{\mu}_{\nu} ({\tilde \omega }) 
\frac{\sin((\mu-\nu)\pi)}{\pi}
\frac{{\rm e}^{-i(\mu+\nu)\pi} 
(1-1/{\tilde r }^{2})^{\mu/2}
({\tilde \omega }{\tilde r })^{\nu}}
{{\tilde \omega }^{\nu} {\tilde r }^{\nu - \mu} 
({\tilde r }-1)^{\mu/2}({\tilde r }+1)^{\mu/2}}
{\rm Qs}^{\mu}_{-\nu-1}({\tilde r };{\tilde \omega })\,,
\nonumber \\
&S^{\mu (2)}_{\nu} ({\tilde r };{\tilde \omega })
= - \sec (\nu \pi) 
\left[
S^{\mu(1)}_{-\nu-1}({\tilde r};{\tilde \omega}) 
+\sin(\nu \pi) S^{\mu (1)}_{\nu} ({\tilde r};{\tilde \omega}) 
\right]\,,
\end{align}
where the factor $ K^{\mu}_{\nu} ({\tilde \omega }) $ in front of the first function is called 
the spheroidal joining factor 
and its  exact form is given by\footnote{In the reference~\cite{Falloon},
this relation is given 
with missing $(-1)^{k}$ factor in the denominator.},
\begin{align}\label{K_exact_form}
K^{\mu}_{\nu}({\tilde \omega })
        ={\rm e}^{i \nu \pi}2^{-2\nu-1} \Gamma(\nu-\mu+1) 
        \frac{{\tilde \omega }^{\nu}}{A^{-\mu}_{\nu}({\tilde \omega })}
        \frac{
        \displaystyle{
        \sum_{k=0}^{\infty} \frac{(-1)^{k} a^{-\mu}_{\nu,-2k}({\tilde \omega })
        }{ \Gamma(-k+\nu+\frac{3}{2})k! }
        }
        }{
        \displaystyle{
        \sum_{k=0}^{\infty} \frac{(-1)^{k} a^{\mu}_{\nu,2k}({\tilde \omega })}{ 
        \Gamma(-k-\nu+\frac{1}{2})k! }
        }
        }\,.
\end{align}
The symmetry relations of the aforementioned the second kind angular spheroidal function of type II, $\text{Qs}^{\mu}_{\nu}({\tilde r };{\tilde \omega })$, and those of the first kind angular spheroidal function of type II, 
$\text{Ps}^{\mu}_{\nu}({\tilde r};{\tilde \omega})$, are given as
\begin{align}
\label{QsPsSym}
&{\rm Qs}_{-\nu-1}^{\mu}({\tilde r};{\tilde \omega})
        = \csc((\mu -\nu) \pi)
        \left[ 
        \pi e^{i\mu\pi} \cos( \nu\pi ) 
        {\rm Ps}^{\mu}_{\nu}({\tilde r };{\tilde \omega })
        - \sin(( \mu+\nu)\pi) 
          {\rm Qs}^{\mu}_{\nu}({\tilde r };{\tilde \omega }) 
        \right], 
        \nonumber \\
&{\rm Ps}^{-\mu}_{\nu} ({\tilde r };{\tilde \omega })
        =\frac{\Gamma(\nu-\mu+1)}{\Gamma (\nu+\mu+1)}
        \left[
        {\rm Ps}^{\mu}_{\nu} ({\tilde r};{\tilde \omega}) 
        - \frac{2}{\pi}e^{-i\mu\pi}\sin(\mu \pi) 
          {\rm Qs}^{\mu}_{\nu} ({\tilde r };{\tilde \omega }) 
        \right],
\end{align}
where $\text{Ps}^{\mu}_{\nu}({\tilde r};{\tilde \omega})$ is expanded in terms of the associated Legendre functions of type II,
$\mathfrak{P}^{\mu}_{\nu}({\tilde r};{\tilde \omega})$. With the help of~\eqref{K_exact_form} and~\eqref{QsPsSym}, we write down the full solutions as a linear combination 
of the first kind angular spheroidal functions of type II 
with different parameters, 
${\rm Ps}^{\mu}_{\nu}({\tilde r };{\tilde \omega })$ and 
${\rm Ps}^{-\mu}_{\nu}({\tilde r };{\tilde \omega })$ as
\begin{align}\label{Sol_NH}
R({\tilde r }) 
        = C_{+} {\rm Ps}^{\mu}_{\nu}({\tilde r };{\tilde \omega })
        +C_{-} {\rm Ps}^{-\mu}_{\nu}({\tilde r };{\tilde \omega })\,, 
\end{align}
where the coefficients $C_{+}$ and $C_{-}$ are given by,
\begin{align}
\label{C+-}
&C_{+}
        = \frac{1}{2}  (1-i\tan (\nu \pi)) 
        \frac{1}{\sin(\mu \pi)}\times
        \nonumber\\
&~~~~~~~\times
        \left[ 
        \sin ((\mu-\nu)\pi)
        (C_{1} \cos ( \nu \pi )-C_{2} \sin (\nu \pi)) 
        K^{\mu}_{\nu}({\tilde \omega })
        - C_{2} e^{2 i \nu \pi } \sin((\mu+\nu)\pi)
        K^{\mu}_{-\nu-1}({\tilde \omega })
        \right] ,
        \nonumber \\
&C_{-} 
        = \frac{1}{2} (1-i\tan (\nu\pi))
        \frac{\sin((\mu+\nu)\pi)}{\sin (\mu \pi)} 
        \frac{\Gamma(\nu+\mu+1)}{\Gamma(\nu-\mu+1)} \times 
        \nonumber \\
&~~~~~~~\times 
        \left[
        (C_{1} \cos (\nu\pi)-C_{2} \sin (\nu\pi))
        K^{\mu}_{\nu}({\tilde \omega }) 
        + C_{2} e^{2 i \nu \pi} K^{\mu}_{-\nu-1}({\tilde \omega }) 
        \right]\,.
\end{align}
With this new form of the full solution, Eq.~(\ref{Sol_NH}), we can directly read the ingoing or outgoing solutions as follows. By definition, the two solutions ${\rm Ps}^{\pm\mu}_{\nu}({\tilde r };{\tilde \omega })$ are written in terms of the first kind associated Legendre functions of type I, $\mathfrak{P}$'s, 
\begin{align}
\label{Ps}
{\rm Ps}^{\pm \mu}_{\nu}({\tilde r };{\tilde \omega })&=\sum_{k=-\infty}^{\infty}(-1)^{k} a^{\pm\mu}_{\nu,2k}({\tilde \omega })\mathfrak{P}^{\pm\mu}_{\nu+2k}({\tilde r})\nonumber \\
&=\frac{1}{\Gamma(1\mp\mu)} 
\frac{({\tilde r}+1)^{\pm\mu/2}}{({\tilde r}-1)^{\pm\mu/2}} \nonumber\\
&~~~\times 
\sum_{k=-\infty}^{\infty} 
(-1)^{k} a^{\pm\mu}_{\nu,2k}({\tilde \omega })
{}_{2}F_{1}
\left( 
-\nu-2k, \nu+2k+1; 1\mp\mu; \frac{1-{\tilde r}}{2}
\right)
,
\end{align}
where ${}_{2}F_{1}(a,b;c;z)$ is Gaussian hypergeometric function. Notice that, in the near horizon limit of ${\tilde r}\rightarrow1$, the series part of~\eqref{Ps} converges to the normalization factors $A^{\pm\mu}_{\nu}({\tilde \omega})$ by its definition Eq.~(\ref{Norm_factor}). Then, the overall factor in front of the series determines the leading behavior of the only remaining dominant terms read 
the ingoing and outgoing asymptotic solutions. Subsequently the coefficients $C_{+}$ and $C_{-}$ in~\eqref{Sol_NH}, specifically~\eqref{C+-}, are identified as those of ingoing and outgoing solutions near the horizon, respectively. By setting the outgoing wave coefficient zero, $C_{-}=0$, we find the ingoing boundary condition,
\begin{align}
\label{C1C2ratio}
\frac{C_{1}}{C_{2}} 
        =\frac{1}{\cos(\pi \nu)} 
        \left[ 
        \sin(\pi \nu) -e^{2 i \nu \pi}
        \frac{ K_{-\nu-1}^{\mu}({\tilde \omega }) }
             { K_{\nu}^{\mu}({\tilde \omega }) } 
        \right]\,.
\end{align}
According to the fixation of the ingoing boundary condition, the ratio of the two spheroidal joining factors, ${K_{-\nu-1}^{\mu}({\tilde \omega }) }/{ K_{\nu}^{\mu}({\tilde \omega }) }$, determines the ratio of two coefficients, $C_{1}/C_{2}$, in the full solution~\eqref{FullSol_SS} up to an overall constant. Note that the ratio $C_{1}/C_{2}$ in~\eqref{C1C2ratio} will be applied only for integral $\nu$'s in the low frequency limit, where the denominator $\cos (\pi \nu)$ becomes $\pm1$. Now that we have the ingoing boundary condition of the 5D Teukolsky equation for brane scalar fields, we are ready to obtain its greybody factors.
\vspace{12pt}

\section{Greybody factors for scalar fields}
\label{Sec4}

The greybody factor for black holes, $\Gamma$, is the correction factor
to the Hawking radiation which is described as a black body radiation. 
This correction factor is calculated 
from the absorption probability of the incoming wave. 
For brane scalar fields, 
the greybody factor $\Gamma$ is given in terms of the ratio 
between the coefficients of ingoing and outgoing far field solutions~\cite{Page}, 
\begin{align}\label{GreyDef}
\Gamma=1-\left| 
         \frac{C_{\rm out}^{\infty}}{C_{\rm in}^{\infty}} 
         \right|^{2}~.
\end{align}
The coefficients of the far field solutions, $C_{\text{in}}^{\infty }$ and $C_{\text{out}}^{\infty }$, are computed by comparing with the asymptotic behavior of the radial spheroidal function in Eq.~(\ref{SS_jy_sc}),
\begin{align}
R({\tilde r }) 
        &= C_{1}S_{\nu}^{\mu(1)}({\tilde r };{\tilde \omega })
        + C_{2}S_{\nu}^{\mu(2)}({\tilde r };{\tilde \omega }) 
        \nonumber \\
&\rightarrow  
        C_{1} j_{\nu}({\tilde \omega} {\tilde r})
        + C_{2} y_{\nu}({\tilde \omega} {\tilde r})  
        \nonumber \\
&\equiv C_{\text{in}}^{\infty} 
        \left( \frac{1}{r} e^{-i \omega r}\right)
        + C_{\text{out}}^{\infty} 
        \left( \frac{1}{r} e^{ i \omega r}\right)~,
\end{align}
where in the last line $\omega r$ was written as the same value 
${\tilde \omega} {\tilde r}=\omega r$. Thus, $C_{\text{in}}^{\infty}$ and $C_{\text{out}}^{\infty}$ are expressed in terms of the coefficients of the spheroidal functions, $C_{1}$ and $C_{2}$,
\begin{align}
\label{Cinfty}
C_{\text{in}}^{\infty}  
        = \frac{C_{1}+iC_{2}}{2{\tilde \omega}} 
        e^{i\frac{(\nu+1)}{2}\pi},\quad
C_{\text{out}}^{\infty} 
        = \frac{C_{1}-iC_{2}}{2{\tilde \omega}} 
        e^{-i\frac{(\nu+1)}{2}\pi}~.
\end{align}
Substitution of~\eqref{Cinfty} into the greybody factor for brane scalar fields~\eqref{GreyDef} gives
\begin{align}
\Gamma =1 - \left\vert 
            -\frac{1+i( C_{1}/C_{2})}
                  {1-i (C_{1}/C_{2})} e^{-i(\nu+1)\pi} 
            \right\vert^{2}~.
\end{align}
Since the ratio $C_{1}/C_{2}$ is fixed by the ingoing boundary condition in Eq.~(\ref{C1C2ratio}), we obtain the greybody factor in terms of the ratio between the spheroidal joining factors,
\begin{align}
\Gamma=1-\left\vert 
         \frac{ie^{i\nu\pi}+e^{2i\nu\pi }
               (K_{-\nu-1}^{\mu}({\tilde \omega })/K_{\nu}^{\mu}(\tilde \omega))}
              {-i+e^{3i\nu\pi} 
               (K_{-\nu-1}^{\mu}({\tilde \omega})/K_{\nu}^{\mu}(\tilde \omega))} 
         \right\vert^{2}~.
\end{align}
As a consistency check, low frequency limit is taken for the obtained solutions and greybody factors, and it is compared
to the known approximate results in~\cite{Ida:2002ez}, whose near-horizon asymptotic solution for brane scalar fields is
\begin{align}
\label{Sol_NH_IOPI}
R_{{\rm NH}}
        =&C'_{1}
        \left( \frac{{\tilde r}-1}{2} \right)^{-\frac{\mu}{2}} 
        \left( \frac{{\tilde r}+1}{2} \right)^{\frac{\mu}{2}}
        {}_{2}F_{1}
        \left( -l,l+1,1-\mu, \frac{1-{\tilde r}}{2} \right)
        \nonumber \\
&+C'_{2}\left( \frac{{\tilde r}-1}{2} \right)^{\frac{\mu}{2}} 
        \left( \frac{{\tilde r}+1}{2} \right)^{\frac{\mu}{2}} 
        {}_{2}F_{1}
        \left( -l+\mu,l+\mu+1,1+\mu, \frac{1-{\tilde r}}{2} \right)\,.
\end{align}
First, in the limit of low energy $\tilde{\omega}\rightarrow 0$, the ratio of two coefficients in the definition of the radial spheroidal function in Eq.~\eqref{Sdef} behaves~\cite{Falloon}
\begin{align}\label{a_k_Ratio}
\frac{a^{\mu}_{\nu,\pm k}(\tilde{\omega})}{a^{\mu}_{\nu,0}(\tilde{\omega})} \sim\mathcal{O}(\tilde{\omega}^{2k}),
~~~~(k=0,1,2,\cdots),
\end{align}
and thus the $a^{\mu}_{\nu,0}(\tilde{\omega})$ term is dominant. Second, the $\nu$ of the solution $S^{\mu(1)}_{\nu}$ or $S^{\mu(2)}_{\nu}$ in~\eqref{FullSol_SS} becomes the integer parameter $l$ of angular eigenvalues~\eqref{A} by the asymptotic property of the spheroidal eigenvalue~\cite{Falloon}. In this limit, with the help of a property of the hypergeometric function, ${}_{2}F_{1} (a,b,c,z)=z^{1-c}{}_{2}F_{1}(1+a-c,1+b-c,2-c,z)$ for non-integral $c$ and a symmetry of the first kind of the angular spheroidal function of type II, ${\rm Ps}_{\nu}^{\mu}({\tilde r };{\tilde \omega }) = {\rm Ps}_{-\nu-1}^{\mu}({\tilde r };{\tilde \omega })$, the full solution in Eq.~\eqref{Sol_NH} turns out to be the same as the approximate solution in Eq.~\eqref{Sol_NH_IOPI} up to a normalization constant.\\

\begin{figure}[H]
\centering
\includegraphics[scale=0.8]{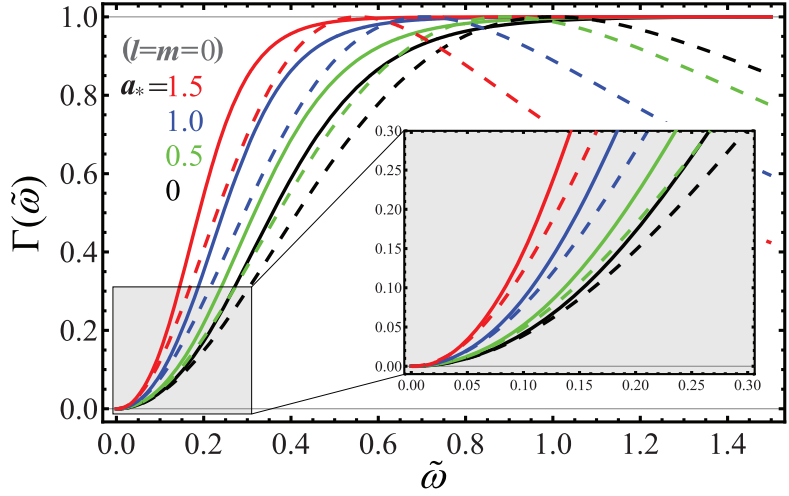}
% Here is how to import EPS art
\caption{\small Greybody factor curves of $l=m=0$ modes are compared for various rotation parameters $a_{*}$: Solid lines denote the graphes of exact solutions and dashed lines those of approximate solutions in the low frequency limit.}
\label{fig:lm0}
\end{figure}

\begin{figure}[H]
\centering
\includegraphics[scale=0.8]{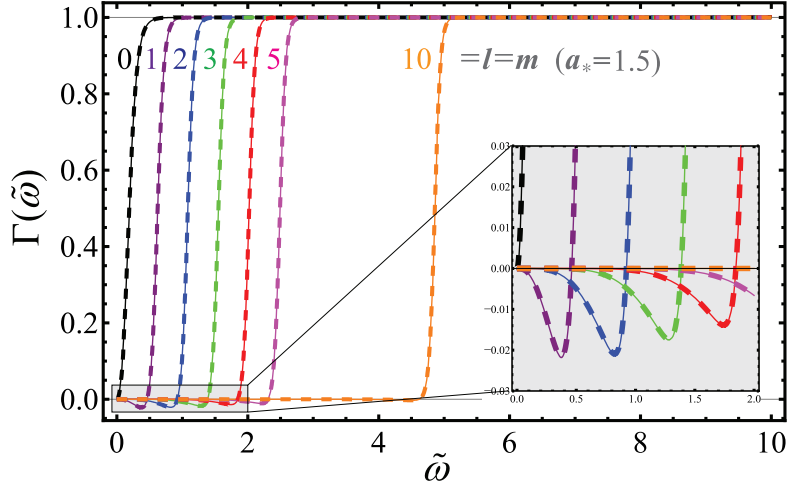}
% Here is how to import EPS art
\caption{\small Greybody factor curves of exact solutions (thin solid lines) and those of numerical results (dashed lines) are compared for $l=m$ modes with $a_{*}=a_{* {\rm max}}$.}
\label{fig:lme}
\end{figure}

In the low frequency limit, the greybody factor is simplified as
\begin{align}\label{Gamma_l}
\Gamma=1-\left\vert 
         \frac{1+i( C_{1}/C_{2})}{1-i (C_{1}/C_{2})} 
         \right\vert^{2}\,,
\end{align}
where the ingoing boundary condition is simplified
\begin{align}
\label{ingoingbc}
\frac{C_{1}}{C_{2}}
        =(-1)^{l+1}
        \frac{ K_{-l-1}^{\mu}({\tilde \omega }) }
             { K_{l}^{\mu}({\tilde \omega }) }\,.
\end{align}
Due to Eq.~(\ref{a_k_Ratio}), the spheroidal joining factor in Eq.~(\ref{K_exact_form}) approaches its asymptotic form
\begin{align}
&K^{\mu}_{\nu}({\tilde \omega })
        \rightarrow 
        e^{il \pi}2^{-2l-1} \Gamma(l-\mu+1)
        \frac{{\tilde \omega }^{l}}
             {A^{-\mu}_{l}({\tilde \omega })}
        \frac{a^{-\mu}_{l,0}({\tilde \omega })\Gamma(-l+\frac{1}{2})}
             {a^{\mu}_{l,0}({\tilde \omega })\Gamma(l+\frac{3}{2})}~.
\end{align}
Using the symmetry relations of the normalizing constant,
$A_{\nu}^{\mu}(\gamma)=A_{-\nu-1}^{\mu}(\gamma)$, 
and spheroidal expansion coefficients, 
$a^{\mu}_{l,0}(\gamma)=a^{-\mu}_{l,0}(\gamma)$, 
the ingoing boundary condition~\eqref{ingoingbc} becomes
\begin{align}\label{CinIOPI}
\frac{C_{1}}{C_{2}}
=(-1)^{l+1} 
\frac{4^{-2l-1}{\tilde \omega}^{2l+1}
\Gamma\left( \frac{1}{2}-l\right)^{2} \Gamma (l-\mu+1)}{\Gamma \left( \frac{3}{2}+l \right)^{2} \Gamma \left( -l-\mu \right)}~,
\end{align}
where~\eqref{ingoingbc} is also used. Straightforward computation of the gamma function in Eq.~\eqref{CinIOPI} leads exactly to the same equation in~\cite{Ida:2002ez}\footnote{In the Ref.~\cite{Ida:2002ez}, the greybody factor was written in terms of $C\equiv -i(C_{1}/C_{2})$.}. In Fig.~\ref{fig:lm0}, the exact solutions and approximate solutions in the low frequency limit are compared for the simplest mode, $(l,m)=(0,0)$. The rotation parameter $a_{*}$ 
varies from zero to the maximum possible value $a_{* {\rm max}}=1.5$.
\vspace{12pt}
In Fig.~\ref{fig:lme}, the greybody factor curves of the exact solutions are also compared to the numerical results~\cite{Ida:2005ax}. We plot these two results for the various $l=m$ modes with the rotation parameter, $a_{*}=a_{* {\rm max}}$, whose graphs show overlaps.
\vspace{12pt}
\vspace{12pt}

\section{Summary and discussions}
\label{Sec5}

In low scale gravity scenarios, higher dimensional rotating black holes are copiously produced in high energy collisions, which may be within the reach of the high energy LHC or future circular collider at CERN or China with $\sqrt{s}=100$ TeV. 
The detection of Hawking radiation from a higher dimensional rotating black hole
may be the most interesting way of proving not only the large or warped extra dimensions~\footnote{In \cite{Orlando:2010kx}, black holes from the compact hyperbolic extra dimensions are discussed}, but also
the quantum phenomenon of gravity~\cite{Cardoso:2012qm}.

To support the experimental searches of such signal and theoretical interest by its own, there have been significant theoretical 
studies on the Hawking radiation for higher dimensional rotating black holes in both the analytical and numerical ways.  
Before this paper, the analytic studies have been successful only in the specific regimes, e.g., low-frequency regime or high-frequency regime.
In the present work, we are successful in finding the  full solution of the generalized Teukolsky equation without assuming any limitation in the frequency domain for the brane scalar modes in 5D black hole background of Myers-Perry solution.  
%For 5D black hole background of Myers-Perry, perturbations of the four-dimensional brane scalar field are taken into account. 
%The time and angular parts of the Teukolsky equation are solved, and then the remaining radial part of the equations reduces to the prolate spheroidal equation with complex parameters. By applying the ingoing boundary conditions, their boundary conditions near the horizon are read
%in terms of the spheroidal joining factors.

The obtained exact result enables us to describe the Hawking radiation in the entire frequency regime and to analyze its characters in detail.
Since the detailed study of greybody factors allow us to have a better understanding of quantum nature of black holes, the study with various approaches will give us some insight into the fundamental properties of black holes. Therefore the exact form of the greybody factor of scalar perturbations for the 5D black holes may provide a good approach for the extra dimension search and development of quantum gravity. 
%The obtained exact result will open new windows for further studies and understandings of the quantum nature of the higher-dimensional black holes in general. When we consider the higher dimensional black holes in the collider, we assume the black hole mass to be large enough to neglect the brane tension at the horizon and small enough to neglect the topology and curvature of the extra dimensions~\cite{Ida:2002ez}. Under this conditions, higher dimensional black holes in the flat background give good approximations. In this framework, our exact solution may shed light eventually on the study on detectable signals from the higher dimensional black holes in LHC.%Once the expression of the boundary condition is obtained, the greybody factor is specifically expressed in terms of the spheroidal joining factors.
%The low frequency limit of our exact result coincides with the approximate result obtained in the same limit~\cite{Ida:2002ez}, and our exact result also matches the previously obtained numerical result~\cite{Ida:2002ez}.

Finally, we comment on the Teukolsky equation for the higher spin fields with $s=1/2,1,2/3$, and $2$.  We have shown that those high spin states are governed by the confluent Heun equation with complex parameters, which still lack of understanding its basic mathematical properties, yet.
%Since this differential equation has not been studied in detail yet, the study on the non-zero spin brane fields requires mathematical analysis prior to physical study. In this manuscript, we focused only to the brane scalar mode of $s=0$ in the 5D black hole background.
Intriguingly, the equation has the same mathematical structure with that of the four-dimensional Kerr black hole~\cite{Mano:1996vt,Mano:1996gn,Sasaki:2003xr}. %which implies intractability of the analysis of the greybody factors from the non-zero spin modes.
Thus, finding analytic solutions in one side would provide the greybody factor on the other side, which is definitely worth more attention for the future.

%\item The obtained exact result will open new windows for further studies and understandings of the quantum nature of the higher-dimensional black holes in general.When we consider the higher dimensional black holes in the collider, we assume the black hole mass to be large enough to neglect the brane tension at the horizon and small enough to neglect the topology and curvature of the extra dimensions~\cite{Ida:2002ez}.Under this conditions, higher dimensional black holes in the flat background give good approximations. In this framework, our exact solution may shed light eventually on the study on detectable signals from the higher dimensional black holes in LHC.

%\item Since the detailed study of greybody factors allow us to have a better understanding of quantum nature of black holes, the study with various approaches will give us some insight into the fundamental properties of black holes. Therefore the exact form of the greybody factor of scalar perturbations for the 5D black holes may provide a good approach for the extra dimension search and development of quantum gravity. 
%\end{itemize}
 
\vspace{12pt}

%\begin{figure}
%\includegraphics[scale=0.8]{}
%\caption{\label{fig:epsart} A figure caption. The figure captions are
%automatically numbered.}
%\end{figure}

%Fig.~\ref{fig:wide} is a figure that is too wide for a single column,
%so instead the \texttt{figure*} environment has been used.
%\begin{figure*}
%\includegraphics{}% Here is how to import EPS art
%\caption{\label{fig:wide}Use the figure* environment to get a wide
%figure that spans the page in \texttt{twocolumn} formatting.}
%\end{figure*}

\acknowledgments

\begin{sloppypar}
Y.-H. Hyun, Y. Kim, and S. C. Park are supported by Basic Science Research Program through the National Research Foundation of Korea (NRF) funded by the Ministry of Science, ICT and Future Planning (Grant No. NRF-2018R1D1A1B07049514,  NRF-2016R1D1A1B03931090, and NRF-2016R1A2B2016112 \& NRF-2018R1A4A1025334, respectively). Y.-H. Hyun is also supported by the Korea
Institute of Science and Technology Information (K-18-L12-C08-S01).
\end{sloppypar}

\appendix

\section{Myers-Perry black holes}
\label{AppA}

The $d$-dimensional flat space metric of $n=[(d-1)/2]~(n\ge 2)$ can take~\cite{Myers},
\begin{align}
&ds^{2}=-{\textrm{d}}t^{2}+\sum_{i=1}^{n}({\textrm{d}}x_{i}^{2}+{\textrm{d}}y^{2}_{i})+{\textrm{d}}z^{2}|_{d_{\text{even}}}
        \nonumber\\
&~~~~=-{\textrm{d}}t^{2}+{\textrm{d}}r^{2}
        +r^{2}\sum_{i=1}^{n}({\textrm{d}}\mu^{2}_{i}+\mu^{2}_{i}{\textrm{d}}\phi^{2}_{i})+r^{2}{\textrm{d}}\alpha^{2}|_{d_{\text{even}}}\,,        
\end{align}
where the paired spatial Cartesian coordinates 
$(x_{i},y_{i})$ in $n$ orthogonal planes
is expressed in terms of the polar coordinates $(\mu_{i}, \phi_{i})$ whose direction cosines are
\begin{align}
&x_{i}=r\mu_{i}\cos \phi_{i},~~~y_{i}=r\mu_{i}\sin \phi_{i},~~z=r \alpha,
        \nonumber\\
&r^{2}=\sum_{i=1}^{n}(x^{2}_{i}+y^{2}_{i}),~~
        \sum_{i=1}^{n}\mu^{2}_{i}+\alpha^{2}|_{d_{\text{even}}}=1.
\end{align} 
With this setup, the solutions of arbitrary rotation 
in every independent rotation plane are given by
\begin{align}
&ds^{2}=-{\textrm{d}}t^{2}
        +\frac{\mu r\cdot (r|_{d_{even}})}{\Pi F}
        \left( {\textrm{d}}t+\sum_{i=1}^{n}a_{i}\mu^{2}_{i}{\textrm{d}}\phi_{i} \right)^{2}
        +\frac{\Pi F}{\Pi-\mu r\cdot (r|_{d_{even}})}{\textrm{d}}r^{2}
        \nonumber\\
&~~~~~~~~
        + r^{2}{\textrm{d}}\alpha^{2} |_{d_{\text{even}}}
        +\sum_{i=1}^{n}(r^{2}+a_{i}^{2})({\textrm{d}}\mu_{i}^{2}
        +\mu^{2}_{i}{\textrm{d}}\phi^{2}_{i})\,,
\end{align}
where the number of independent rotation planes is $n=\left[ \frac{d-1}{2} \right]$ and
\begin{align}
\mu^{2}_{i}+\alpha^{2}=1,~~
        F=1-\sum_{i=1}^{n}\frac{a^{2}_{i}\mu^{2}_{i}}{r^{2}+a^{2}_{i}},~~
        \Pi=\prod_{i=1}^{n}(r^{2}+a^{2}_{i})\,.    
\end{align}
In the obtained metric, ($r|_{d_{\text{even}}},~d\alpha^{2}|_{d_{\text{even}}}$) reduce to $(1,0)$ for $d$ odd dimensions. 
By examining the asymptotic structure of these metrics,
the $(n+1)$ free parameters, $\mu$ and $a_{i}$, determine respectively the mass and angular momenta carried by the black hole,
\begin{align}
M=\frac{(d-2)\Omega_{d-2}}{16\pi G}\mu,~~
        J^{y_{i}x_{i}}=\frac{\Omega_{d-2}}{8\pi G}\mu a_{i}
        =\frac{2}{d-2}Ma_{i}\,.
\end{align}
\vspace{12pt}

\section{Teukolsky equations on brane}
\label{AppB}

We consider a scalar field in background of the 5D Myers-Perry black holes. The scalar field propagates on the four-dimensional brane and obeys the linear equation coupled to the background gravity. In this section, by use of separation of the variables and subsequently by solving some part of the obtained ordinary differential equations, we show that the resultant decoupled ordinary differential equations become uniquely the generalized Teukolsky equations of $s=0$. \\

We begin our discussion with the introduction of a real scalar field $\phi(x)$ whose dynamics is governed by the Klein-Gordon equation,
\begin{align}
\label{KGeq'}
\nabla^{2}\phi=\frac{1}{\sqrt{-g}}\partial_{\mu}(\sqrt{-g}\,g^{\mu\nu}
\partial_{\nu}\phi)=0
~.
\end{align}
This scalar field $\phi$ propagates on the four-dimensional brane in the five-dimensional bulk formed by the Myers-Perry black hole~\cite{Myers} whose metric is expressed in the Boyer-Lindquist coordinates as
\begin{align}\label{5Dmetric}
ds^2 =&\left(
        1-\frac{{\tilde \mu}}{\Sigma(r,\theta)}
        \right){\rm d}t^2
        +2a\sin^{2}\theta
        \frac{{\tilde \mu}}{\Sigma(r,\theta)}{\rm d}t{\rm d}\phi
        -\sin^{2}\theta
        \left(
        r^2 +a^2 +a^2 \sin^2 \theta \frac{{\tilde \mu}}{\Sigma(r,\theta)}
        \right) 
        {\rm d}\phi^2
        \nonumber\\
&-\frac{\Sigma(r,\theta)}{\Delta(r)}{\rm d}r^2
        -\Sigma(r,\theta) {\rm d}\theta^2
        -r^2 \cos^2 \theta {\rm d}\psi^{2} ,
\end{align}
where the functions in the metric have
\begingroup
\allowdisplaybreaks
\begin{align}
\Sigma(r,\theta) &\,= r^{2} + a^{2} \cos^{2} \theta
\,,
\\
\Delta(r) &\,= r^2 + a^2 - {\tilde \mu} \equiv r^{2}-r_{{\rm h}}^{2}
\,.
\end{align}
\endgroup 
For the Myers-Perry black hole, see Appendix~\ref{AppA}.
The parameters ${\tilde \mu}$ and $a$
are a normalized mass and a normalized angular momentum, respectively. 
From $\Delta(r_{\rm h})=0$, we read an event horizon at $r=r_{{\rm h}}$. Thus the induced metric for the brane scalar field in four dimensions is obtained by neglecting $\psi$ dependence, ${\textrm{d}}\psi=0$ in \eqref{5Dmetric}, and, in this background gravity, the brane scalar field follows the linear equation from \eqref{KGeq'},
\begin{align}
\label{KG_in_five'}
&\bigg[
\left( 1+\frac{\mu(r^{2}+a^{2})}{\Delta\Sigma} \right)
\frac{\partial ^{2}}{\partial t^{2}}
-\frac{1}{\Sigma}\frac{\partial}{\partial r}
\Delta\frac{\partial}{\partial r}
-\frac{1}{\Sigma\sin\theta}\frac{\partial}{\partial \theta}
\sin\theta\frac{\partial }{\partial \theta}
\nonumber\\
&~~
-\frac{1}{\Delta\sin^{2}\theta}\left( 1-\frac{\mu}{\Sigma} \right)
\frac{\partial^{2}}{\partial \varphi^{2}}
+\frac{2a\mu}{\Delta\Sigma}
\frac{\partial ^{2}}{\partial t\partial \varphi }
\bigg]\phi=0
\,.
\end{align}

Separation of variables $\phi\equiv T(t)R(r)Y(\theta,\varphi)$ to the 
equation~\eqref{KG_in_five'} results in
\begin{align}
\label{T(t)'}
\frac{{\rm d}^{2} T(t)}{{\rm d} t^{2}}
+f(r,\theta,\varphi)\frac{{\rm d}T(t)}{{\rm d} t}
+g(r,\theta,\varphi)T(t)
=0
\,,
\end{align}
where $f$ and $g$ are time-independent but involve both the spatial variables $(r,\theta,\varphi)$ and their derivatives as
\begin{align}
&
f=
\left( 1+\frac{\mu(r^{2}+a^{2})}{\Delta\Sigma} \right)^{-1}
\frac{2a\mu}{\Delta\Sigma}\frac{1}{Y(\theta,\varphi)}
\frac{\partial Y(\theta,\varphi)}{\partial \varphi}
\,,
\label{f'}
\\
&
g=
-\left( 1+\frac{\mu(r^{2}+a^{2})}{\Delta\Sigma} \right)^{-1}
\bigg[
\frac{1}{\Sigma}\frac{1}{R(r)}\frac{{\rm d}  }{{\rm d} r}
\left(\Delta\frac{{\rm d}  R(r)}{{\rm d} r}\right)
+\frac{1}{\Sigma\sin\theta}\frac{1}{Y(\theta,\varphi)}
\frac{\partial}{\partial \theta}
\left(\sin\theta\frac{\partial Y(\theta,\varphi)}{\partial \theta}\right)
\nonumber\\
&~~~~~~~~~~~~~~~~~~~~~~~~~~~~~~~~~~~~~
+\frac{1}{\Delta\sin^{2}\theta}\left(1-\frac{\mu}{\Sigma}\right)
\frac{1}{Y(\theta,\varphi)} 
\frac{\partial^{2}Y(\theta,\varphi)}{\partial\varphi^{2}}
\bigg]
\,.
\label{g'}
\end{align} 
The time-dependent part $T(t)$ of the linear equation~\eqref{T(t)'} is solved by the general solution,
\begin{align}
\label{T(t)sol'}
T(t)=
C_{+}e^{-i\omega_{+}t}+C_{-}e^{-i\omega_{-}t}
%C_{1}e^{-\frac{1}{2}(f+\sqrt{f^{2}-4g})t}
%+
%C_{2}e^{-\frac{1}{2}(f-\sqrt{f^{2}-4g})t}
\,,
\end{align}
where $\omega_{\pm}=-\frac{i}{2}(f\pm\sqrt{f^{2}-4g}\,)$. Since $T$ depends only on time $t$ for both non-zero coefficients $C_{\pm}$, constancy of the two frequencies $\omega_{\pm}$ requires both $f$ and $g$ to be constants independent of other variables 
$(r,\theta,\varphi)$. Further separation of variables, $Y(\theta,\varphi)\equiv S(\theta) \Phi(\varphi)$, reduces the equation \eqref{f'} to the equation of $\Phi$,
\begin{align}
\label{f_const_eq'}
\frac{\partial \Phi(\varphi)}{\partial \varphi}
=
i\frac{\omega_{+}+\omega_{-}}{2a \mu}\left[ \Delta \Sigma+\mu(r^{2}+a^{2}) \right]
\Phi(\varphi)
\,,
\end{align}
where $f$ is replaced by $\omega_{\pm}$. Since the functions in front of $\Phi(\varphi)$ in the right-hand side does not depend on the variable 
$\varphi$, its solution up to an overall constant is given with a constant $m$ by
\begin{align}
\label{Phi_sol'}
\Phi(\varphi)= e^{im \varphi}
\,,~~~~(m\in \mathbb{Z})\,
\end{align}
which leads to  a constraint,
\begin{align}
\frac{\Delta\Sigma+\mu(r^{2}+a^{2})}{2a\mu}(\omega_{+}+\omega_{-})=m\,.
\end{align}
Since this algebraic equation can become meaningful only for $m=0$ and $\omega_{+}+\omega_{-}=0$, the constant $f$~in~\eqref{f'} is given by zero and thus, from~\eqref{T(t)'} $\varphi$-dependence should disappear, $Y(\theta,\varphi)=S(\theta)$.

Let us consider another case of $C_{+}=0$ or $C_{-}=0$. Then the solutio has $T(t)= e^{-i \omega t}$ with constant $\omega$ of $2i \omega=f\pm\sqrt{f^{2}-4g}$. Further separation of variables, $Y(\theta,\varphi)\equiv S(\theta)\Phi(\varphi)$, makes the equation be
\begin{align}
\frac{{\textrm{d}}^{2}\Phi(\varphi)}{{\textrm{d}} \varphi^{2}}+p(r,\theta)\frac{{\textrm{d}} \Phi(\varphi)}{{\textrm{d}} \varphi}+q(r,\theta)\Phi(\varphi)=0\,,
\end{align}
where the two functions $p(r,\theta)$ and $q(r,\theta)$ are
\begin{align}
&p(r,\theta)=\frac{2a \mu \omega \sin ^{2}\theta}{\Sigma-\mu}\,,\nonumber\\
&q(r,\theta)=\frac{\Delta \sin ^{2}\theta}{\Sigma-\mu}\bigg[ \frac{1}{R(r)}\frac{{\textrm{d}} }{{\textrm{d}} r}\left( \Delta \frac{{\textrm{d}} R(r)}{{\textrm{d}} r} \right)+\frac{1}{\sin\theta}\frac{1}{S(\theta)}\frac{{\rm d}  }{{\rm d} \theta}\left(\sin\theta\frac{{\rm d} }{{\rm d} \theta}S(\theta)\right)\nonumber\\
&~~~~~~~~~~~~~~~~~~~~~~~~-\omega^{2}\left( \Sigma+\frac{\mu(r^{2}+a^{2})}{\Delta} \right)\bigg]\,.
\end{align}
In the same manner, the solution of $\Phi(\varphi)$ can takes the form,
\begin{align}
\label{T(t)sol'}
\Phi(\varphi)=
C'_{+}e^{im_{+}\varphi}+C'_{-}e^{im_{-}\varphi}
%C_{1}e^{-\frac{1}{2}(f+\sqrt{f^{2}-4g})t}
%+
%C_{2}e^{-\frac{1}{2}(f-\sqrt{f^{2}-4g})t}
\,,
\end{align}
where $m_{\pm}=\frac{i}{2}(p\pm\sqrt{p^{2}-4q}\,)$. The coordinate dependence of $p(r,\theta)$ again restricts the coefficients as $C'_{+}=0$ or $C'_{-}=0$, that leads to~\eqref{Phi_sol'} with an arbitrary integer $m$. The straightforward calculation results in the generalized Teukolsky equation for brane scalar fields,
\begin{align}
\label{RS}
&\frac{{\rm d}  }{{\rm d}r}\left(\Delta\frac{{\rm d}  }{{\rm d} r}R(r)\right)+\bigg[
        \frac{K^{2}}{\Delta}
        +2ma\omega-a^{2}\omega^{2}-A
        \bigg] R(r)=0\,,\nonumber\\
&\frac{1}{\sin\theta}\frac{{\rm d}  }{{\rm d} \theta}\left(\sin\theta\frac{{\rm d} }{{\rm d} \theta}S(\theta)\right)+[(a \omega \cos\theta)^{2}
        -(m\csc\theta)^{2}\textcolor{magenta}{}+A]S(\theta)=0\,.
\end{align}
Note that the previous case is nothing but the spacial case of $m=0$ case in this solution.
\vspace{12pt}

\section{Spheroidal equation}
\label{AppC}

The spheroidal equation takes the form,
\begin{align}
\label{spheroidaleq}
&\frac{{\textrm{d}}}{{\textrm{d}} z} (1-z^{2}) 
        \frac{{\textrm{d}}}{{\textrm{d}}z} w(z)+\Big[
        \gamma^{2}(1-z^{2})
        + \lambda 
        - \frac{\mu^{2}}{1-z^{2}}
        \Big]w(z)=0\,,
\end{align}
where the parameters $\mu$, $\nu$, and $\gamma$ are called 
the order, degree, and size parameters, respectively. According to the parameters, the spheroidal equation~\eqref{spheroidaleq} is classified by four equations as
\begin{center}
\begin{tabular}{|c|c|}\hline
parameter & name of spheroidal equation \\\hline\hline
$|z|<1,~\gamma\in \mathbb{R}~$ & angular prolate spheroidal equation \\\hline
$|z|<1,~\gamma\in i\mathbb{R}$ & ~angular oblate spheroidal equation \\\hline

$|z|>1,~\gamma\in \mathbb{R}~$ & ~~radial prolate spheroidal equation \\\hline

$|z|>1,~\gamma\in i \mathbb{R}$ & ~~~radial oblate spheroidal equation \\\hline
\end{tabular}
\end{center}
 
For the real (purely imaginary) size parameter $\gamma$, 
the equation is called the prolate (oblate) spheroidal equation.
The constant $\lambda_{\nu}^{\mu}(\gamma)$ 
is called spheroidal eigenvalue 
and is determined as the minimal solution~\cite{Falloon}. 
When $\gamma=0$ and $|z|<1$, the equation~\eqref{spheroidaleq} reduces to Legendre differential equation. When $f(z)=(1-1/z^{2})^{\mu/2}g(\gamma z)$, $\gamma \rightarrow 0$, and $|z|>1$, it reduces to the spherical Bessel equation. The solutions to this equation for the separated domain
is defined as,
\begin{center}
\begin{tabular}{|c|}\hline
$~~~~~~~~~~~~~~~~~~~~~~~~~
        |z|<1~~~~~~~~~~~~~~~~~~~~~~~~~~~~~~
        1< |z|~~~~~~~~~~~~$ \\\hline\hline
$~~~~\text{first kind : }ps~(P)
        \xrightarrow{~~~~~~~~~~~~~~~~~~~~~~~~~}
        Ps~(\mathfrak{P})
        ~~~~~~~~~~~S^{(1)}(j)$ \\
$\text{second kind : }
        qs~(Q)\xrightarrow{\text{\scriptsize ~~~analytic continuation~  }}
        Qs~(\mathfrak{Q})
        ~~~~~~~~~~S^{(2)} (y)$ 
        \\\hline
\small{~~~~~~~~~~~~angular (type I)~~~~~~~~~~~~~~~~~~~~~~~
        angular (type II)~
        radial solutions}\\\hline
\end{tabular}
\end{center}
where the expansion functions in the parenthesis are $P,Q,\mathfrak{P},\mathfrak{Q}$ 
(associated Legendre functions) and $j,y$ (spherical Bessel functions).\\

The angular and radial solutions in the table are expanded as
\begin{align}
&F^{\mu}_{\nu}(z;\gamma)
        =\sum_{k=-\infty}^{\infty}
        (-1)^{k}a^{\mu}_{\nu,2k}(\gamma)f^{\mu}_{\nu+2k}(z),
        ~(F=ps,qs,Ps,Qs,~f=P,Q,\mathfrak{P},\mathfrak{Q})\,,
        \nonumber\\
&S^{\mu(i)}_{\nu}(z;\gamma)
        =\frac{(1-1/z^{2})^{\mu/2}}{A^{-\mu}_{\nu}(\gamma)}
        \sum_{k=-\infty}^{\infty}a^{-\mu}_{\nu,2k}
        f_{\nu+2k}(\gamma z),~(f=j,y)\,,        
\end{align}
where the coefficient $a^{\mu}_{\nu,2k}$ and 
the normalization factor $A^{\mu}_{\nu}(\gamma)$ satisfy\footnote{
Typos in the recurrence relation in the ref.~\cite{Falloon} has been corrected
in the $a^{\mu}_{\nu,2k}$ indices and numerical factors 
in the denominator of the 
$B^{\mu}_{\nu,2k}$ relation.}
\begin{align}
&A^{\mu}_{\nu,2k}(\gamma)a^{\mu}_{\nu,2k-2}(\gamma)
        +(B^{\mu}_{\nu,2k}(\gamma)-\lambda)a^{\mu}_{\nu,2k}(\gamma)
        +C^{\mu}_{\nu,2k}(\gamma)a^{\mu}_{\nu,2k+2}(\gamma)=0\,,
        \nonumber\\
&\begin{cases}
\displaystyle{A^{\mu}_{\nu,2k}(\gamma)=-\gamma^{2}
        \frac{(\nu-\mu+2k-1)(\nu-\mu+2k)}{(2\nu+4k-3)(2\nu+4k-1)}} \\
\displaystyle{B^{\mu}_{\nu,2k}(\gamma)=(\nu+2k)(\nu+2k+1)
        -2\gamma^{2}
        \frac{(\nu+2k)(\nu+2k+1)+\mu^{2}-1}{(2\nu+4k-1)(2\nu+4k+3)}} \\
\displaystyle{C^{\mu}_{\nu,2k}(\gamma)=-\gamma^{2}
        \frac{(\nu+\mu+2k+1)(\nu+\mu+2k+2)}{(2\nu+4k+3)(2\nu+4k+5)}} \\
\end{cases}\,,\nonumber\\
&A^{\mu}_{\nu}(\gamma)
        =\sum_{k=-\infty}^{\infty}(-1)^{k}a^{\mu}_{\nu,2k}(\gamma)\,.     
\end{align}
The series expansions of the angular and radial solutions are convergent 
only when the coefficients $a^{\mu}_{\nu,2k}(\gamma)$ form a minimal solution to the recurrence relation. To be specific,
 a solution with the property that 
$a^{\mu}_{\nu,2k\pm2}(\gamma)/a^{\mu}_{\nu,2k}(\gamma)\rightarrow0$
as $k\rightarrow \pm \infty$ converges for all values of $z$.
There is a countably infinite set of values of $\lambda$,
which correspond to minimal solutions. The spheroidal eigenvalue $\lambda^{\mu}_{\nu}(\gamma)$
is defined as a function of $\mu, \nu$, and $\gamma$
by choosing the minimal $\lambda$ value which reduces to $\nu(\nu+1)$
continuously in the limit of $\gamma\rightarrow0$ along the line $(0,\gamma)$~\cite{Falloon}. 
The expansion of the radial solution is absolutely convergent for $|z|>1$,
and the normalization factor $A^{\mu}_{\nu}(\gamma)$ is chosen
so that the radial function shows the following behaviour 
for $\gamma z\rightarrow\infty$,
\begin{align}
&S^{\mu(1)}_{\nu}\rightarrow\frac{1}{\gamma z}
        \sin \left( \gamma z-\frac{\nu \pi}{2} \right)\,,~~~~S^{\mu(2)}_{\nu}\rightarrow-\frac{1}{\gamma z}
        \cos \left( \gamma z-\frac{\nu \pi}{2} \right)\,.
\end{align}
The functions $S^{\mu(1,2)}_{\nu}(z;\gamma)$ 
have branch cuts in the complex $z$-plane. One branch cut lies along the semi-infinite line which begins at the point $z = 0$ and passes $z=-1/\gamma$
for non-integer $\nu$, and the other branch cut does on the interval $(-1,1)$ for non-integer $\mu/2$.

\end{document}